\documentclass[reprint,prx,twocolumn,superscriptaddress,floatfix,showkeys,nolinenumbers,longbibliography]{revtex4-2}
\usepackage{amsmath}
\usepackage{booktabs}
\usepackage{color}
\usepackage{graphicx}
\usepackage{dcolumn}
\usepackage{bm}
\usepackage{array}
\usepackage{multirow}
\usepackage{fix-cm}
\usepackage{pbox}
\usepackage{amssymb}
\usepackage{amsmath}
\usepackage{wasysym}
\usepackage{float}
\usepackage{sidecap}
\usepackage[normalem]{ulem}
\usepackage{epstopdf}
\usepackage[skins]{tcolorbox}
\usepackage{multirow}
\usepackage{ifthen}

\begin{document}

\title{Understanding the phase behavior of a proto-biomembrane}
\author{John F. Nagle}
\email[E-mail: ]{nagle@cmu.edu}
\affiliation{$^1$Department of Physics, Carnegie Mellon University, Pittsburgh, PA 15213, USA}

\date{\today}


\begin{abstract}
The rich thermotropic behavior of lipid bilayers is addressed using phenomenological theory informed by many experiments. The most recent experiment not yet addressed by theory has shown that the tilt modulus in DMPC lipid bilayers decreases dramatically as the temperature is lowered toward the main transition temperature $T_M$. It is shown that this behavior can be understood by introducing a simple free energy functional for tilt that couples to the area per molecule.  This is combined with a chain melting free energy functional in which the area is the primary order parameter that is the driver of the main transition. Satisfactory agreement with experiment is achieved with values of the model parameters determined by experiments, but the transition is directly into the gel phase. The theory is then extended to include the enigmatic ripple phase by making contact with the most recent experimentally determined ripple structure. 
\end{abstract}



\maketitle


\section{Introduction}
Proto-biomembranes consisting of lipid bilayers have fascinating thermodynamic phase behavior even when an artificial membrane is formed with only one of the many lipids found in organisms.  When immersed in water, phosphocholine (PC) lipids that have two saturated hydrocarbon chains, both of chain length $n$ (for $n$ = 14-18), have four phases and three transition temperatures that depend upon the chain length.  
The high temperature phase is often called the fluid phase because the lipids in the two-dimensional membrane are disordered and mobile. It is often identified by the symbol $L_{\alpha}$. (Biophysics literature often calls this the liquid-crystalline phase, although the other phases are also considered liquid crystals in physics.) Most of the membranes in organisms are in a fluid phase.
As temperature is lowered, the lipids become better ordered at the main phase transition temperature $T_M$, but the bilayer is far from crystalline and it takes an enigmatic ripple ($P_{\beta'}$) 
structure~\cite{tardieu1973structure,mcconnell1980rippled,wack1989synchrotron,sun1996structure,zasa1997high,sengupta2003structure,akabori2015structure} which has been a major challenge for physical understanding. 
Further reduction in temperature through the so-called pre-transition or lower transition at $T_L$ takes the bilayers into the misnamed gel ($L_\beta'$) phase which still retains considerable disorder \cite{sun1994order}; skin membranes include gel-like regions \cite{mojumdar2013localization,Huster2023skin}. 
Even further reduction in temperature, while still remaining above the freezing point of water in which the bilayers are immersed, very slowly form a subgel phase ($L_C$) that begins to show signatures of two-dimensional crystallinity which are still not well characterized structurally and likely have no biological importance. 

This paper focuses on the fluid ($F$), ripple ($R$) and gel ($G$) phases and the main and lower transitions of the PC lipid DMPC which has two saturated linear hydrocarbon chains, each with 14 carbons bonded via a glycerol moiety to a PC headgroup.  
It has been widely recognized that the main phase transition of DMPC at $T_M$ = 24.0 $^\circ$C is first order with a latent heat $\Delta{H}$ = 6.5 kcal/mole \cite{blume1983apparent} and discontinuous jumps in structural quantities, notably a 27\% increase in area per molecule from 0.47 nm$^2$ \cite{tristram2002structure} to 0.60 nm$^2$ \cite{kuvcerka2005structure} and a 2.7\% increase in volume \cite{nagle1978lecithin}. 
However, the temperature dependence of the volume above the transition was noted as possibly signifying the existence of a critical point at an experimentally inaccessible point in an extended phase diagram. Although this was a rather small effect, there have also been other suggestions of pseudocriticality from experiments \cite{mitaku1983thermodynamic,hatta1984evidence}.

\begin{figure}
\includegraphics[width=1.0\columnwidth,angle=0]{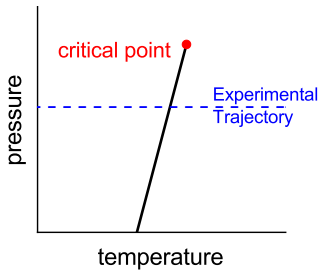}{}
	\caption[]{The solid line shows the locus of a first order transition that ends in a critical point.}
	\label{F1}
\end{figure}

Recently, more dramatic critical-like behavior above the main transition has been observed when studying the temperature dependence of mechanical moduli in DMPC \cite{nagle2017x}.  Theories of the mechanical behavior of membranes originally focused at long length scales where the bending modulus $K_C$ dominates. As the molecular length scale is approached, molecular tilt becomes important in physical studies.  It is a degree of freedom that overcomes an otherwise insurmountable barrier to biological membrane fusion and fission \cite{kozlovsky2002stalk}.  The new finding regards the tilt modulus $K_m$.  Like $K_C$, it is like the stiffness of a spring and its inverse $1/K_m$ is like a compressibility. The tilt modulus decreases by a factor of 3 when $T$ decreases from 40$^\circ$C to the transition at TM, = 24$^\circ$C. This is unlike most stiffness properties that increase with decreasing temperature, but it is what is observed near a critical point.  
Although $K_m$ does not reach zero, which would be an infinite critical compressibility 1/$K_m$, the idea that critical behavior is observable even when the transition is ultimately first order is well understood.  Figure \ref{F1} shows how this occurs in a simple fluid.  When the pressure is constrained, the thermal trajectory may cross the first order phase line, but still lie within a critical region surrounding the critical point where the compressibility becomes large.   Of course, for simple fluids, pressure and temperature can be varied to achieve an experimental trajectory through the critical point, but similar experiments have yet to be found for lipid bilayers.

The pertinent thermodynamic quantities in theories of phase transitions near critical points are a reduced or relative temperature $t$ and an order parameter $\alpha$. 
Of course, lipid molecules are much more complex than the substituents in typical simple fluids and the interaction with water to form bilayers adds another level of complexity.  One should therefore not be surprised that there would be several different order parameters that could interact with each other in interesting ways \cite{goldstein1989structural}. 
This paper addresses this by developing a phenomenological, continuum, Landau-deGennes-like description of the free energy. This follows many previous papers that have developed continuum theories for lipid bilayers \cite{doniach1979thermodynamic,marder1984theory,carlson1987theory,goldstein1988model,chen1995phase,mackintosh1997internal,hansen1998fluid,andelman2008phase,raghu2012phase}.  
While some of these theories have provided connections to the molecular level \cite{doniach1979thermodynamic,carlson1987theory}, generally the continuum models involve phenomenological parameters that do not relate to molecular interaction energies. Nevertheless, continuum-level models can provide insight into the broad features of a system and its phase transitions, more so when the results of the parameterized model agree quantitatively with much experimental data; the model in this paper is compared to more data than previous theories. 

This paper develops free energy functionals for two types of order parameters.  Section \ref{sec2} focuses on the hydrocarbon chains whose conformational change from essentially atraight (all-trans) at low temperature to disordered conformations in the fluid phase; this chain disordering (melting) has long been recognized as the driver of the main transition \cite{nagle1980theory}. This section emphasizes that assuming a conventional free energy functional that works for simple fluids is not necessarily the best choice for the more complex state of lipids in a bilayer. Section \ref{sec3} focuses on molecular tilt to make contact with the new experimental results for the tilt modulus $K_m$.  Section \ref{sec4} shows results obtained from an intermediate theory that combines the free energies functionals from Sections \ref{sec2} and \ref{sec3}.  While this intermediate theory accommodates a good deal of experimental data, including the new data for the temperature dependence of tilt modulus, it only provides a main transition from the fluid phase to a gel phase. Section \ref{sec5} then reviews the heterogeneous structure of the intervening ripple phase.  Earlier theories \cite{doniach1979thermodynamic,marder1984theory,carlson1987theory,goldstein1988model,chen1995phase,mackintosh1997internal,hansen1998fluid,andelman2008phase,raghu2012phase} are followed in Section \ref{sec6} by invoking a term in the free energy functional that depends on this heterogeneity, which then provides both the main and the lower transitions.  While this is not deemed completely satisfactory, as discussed in Section \ref{sec7}, it is suggested that this continuum theory is nevertheless an advance on previous theories.  
 
 \section{Chain Melting Free Energy Functional $F_C$}
\label{sec2}

Conformational disordering of the hydrocarbon chains, i.e. chain melting, is clearly the dominant feature of the main transition \cite{nagle1980theory}. Two likely quantities for the chain melting order parameter are either the difference in the area per molecule or the difference in thickness between the fluid phase and the gel phase.  This is not a major choice because area times thickness is volume and there is only a small percentage volume change at the main transition \cite{nagle1978lecithin}.  Area $A$ is chosen and the order parameter is defined as
\begin{eqnarray}
\alpha = A - A_0.
\label{alpha}
\end{eqnarray}
Here $A_0$ = 0.40 nm$^2$ is twice the cross sectional area of the hydrocarbon chains in the gel phase. It is important to emphasize that $A_0$ is not the surface area per DMPC molecule in the gel phase whose value is $A_G = 0.47$ nm$^2$ \cite{tristram2002structure}; instead, $A_C = A_G$cos($\theta_G$) takes into account that chains tilted by $\theta_G$ = 32$^\circ$ \cite{tristram2002structure} are closer together than the headgroups. This convention assigns $\alpha$ = 0 to the gel phase.  In the fluid phase, disordered chains have no average tilt, so $A$ is then the headgroup area.

\begin{figure}[ht]
\begin{center}
\advance\leftskip-2.5cm
\advance\rightskip-3cm
\vspace{-0.7cm}
\includegraphics[keepaspectratio=true,scale=0.36]{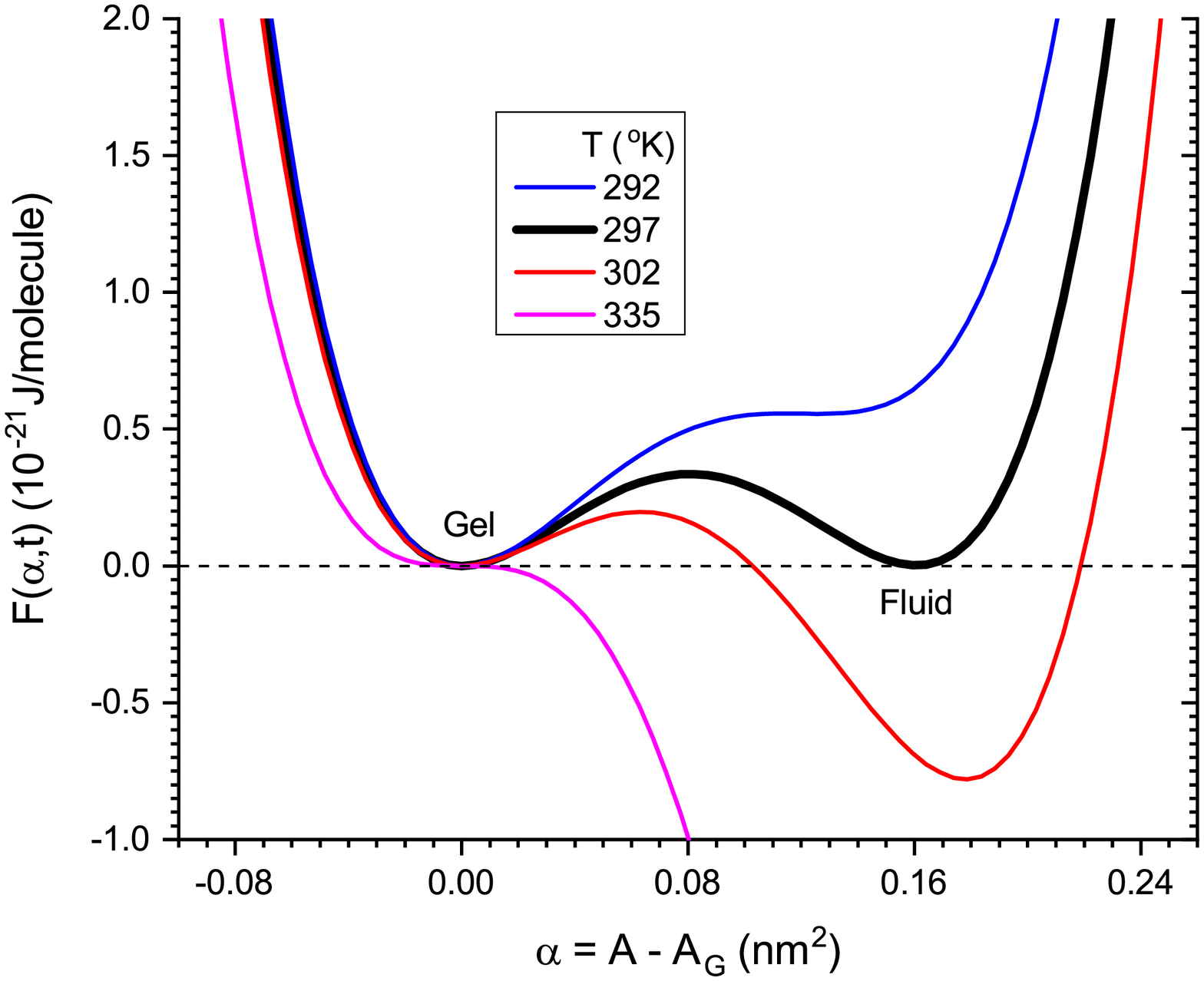}
\vspace{-0.5cm}
	\caption[]{Isotherms for the $\phi^4$ free energy functional in Eq.~(\ref{Fphi4}) with $b_2$=-10.8x10$^{-21}$J/nm$^4$, $b_3$=-7.74x10$^{-18}$J/nm$^6$, $b_4$=3.21x10$^{-17}$J/nm$^8$, and $T_C$=335 K.}
	\label{F2}
\end{center}
\end{figure}

A major choice regards the form of the free energy functional. If one slavishly adopts the conventional form for magnetism or simple fluids, one writes 
\begin{equation}
F_C(\alpha,t)=\frac{1}{2}b_2t\alpha^2+\frac{1}{3}b_3\alpha^3+\frac{1}{4}b_4\alpha^4
\label{Fphi4},
\end{equation}
where $t$ is defined as
\begin{equation}
	t = T  -  T_C . 
 \label{eqt}
\end{equation}
 Negative values of $b_2$ and $b_3$ bring about a first order transition as illustrated in Fig.~\ref{F2}.  The critical point is pushed into a different place in parameter space that is quite likely difficult to achieve in experiments on lipid bilayers.  That is consistent with the suggestion that critical behavior affects the phase transition even though it is ultimately a first order transition \cite{nagle1980theory,goldstein1988model}. 
 
 There is, however, a problem with the model in Eq.(\ref{Fphi4}). The area compressibility modulus $K_A/A$ is the curvature in the isotherms at their minima for a flaccid bilayer with zero surface tension.  Figure \ref{F2} indicates that the curvatures are equal for the gel and fluid phases and this is proven in the Appendix.  Therefore, $K_A$  has only a slightly larger value in the gel phase than in the fluid phase, by the ratio of $A_F/A_G$.  Although the gel phase $K_{AG}$ is relatively poorly determined experimentally, it is clearly much larger than $K_{AF}$ in the fluid phase \cite{needham1988structure,lee2001all} and a simulation gives a ratio $K_{AG}/K_{AF}$ about 4.6 \cite{tjornhammar2014reparameterized}. 

This paper instead chooses a free energy functional form extracted from a microscopic toy model of chain melting \cite{nagle1975critical}.  This toy model emphasized the hard-core, steric, excluded volume interaction between hydrocarbon chains in competition with trans-gauche type conformational disordering. In contrast to the soft interactions of order $kT$ between spins in Ising models, hard-core, excluded volume interactions are essentially either infinite or zero compared to $kT$. Like the two-dimensional Ising model, the statistical mechanics of the toy model were exactly calculable, but with major differences in thermodynamic behavior, even at the qualitative level. In the spirit of free energy functional theory, let us use the lowest order approximation for the equation of state of that model \cite{nagle1975critical} that applies near its critical point that occurs at $T_C$ and chain packing area $A_0$. 
In terms of $t$ in Eq.~(\ref{eqt}) and $\alpha$ in Eq.~(\ref{alpha}), the equation of state for the surface pressure $\pi$ is
 
 \begin{equation}
 \pi= Bt - C(\alpha^2 + 2{\alpha}Dt) + \pi_c	
 \label{eq4}
 \end{equation} 
 
for $\alpha$ greater than 0. In the toy model, the smallest achievable area is $\alpha$ = 0 due to the hard core steric interaction of packing all-trans hydrocarbon chains. For an incompressible chain packing phase there is a minimum area at $A_0$, so $\pi$ at $\alpha$= 0 is not constrained by Eq.~(\ref{eq4}) but can take values up to infinity with no further decrease in $\alpha$.  This is a completely incompressible gel phase, where the incompressibility refers to the chains, not the headgroups which will appear in the next section.  The constant $\pi_c$ in Eq.~(\ref{eq4}) will be chosen to ensure that the experimental trajectory has $\pi$ = 0 corresponding to lipid bilayers that are experimentally flaccid with no tension or pressure.
 
\begin{figure}[ht]
\begin{center}
\advance\leftskip-2.5cm
\advance\rightskip-3cm
\vspace{-0.5cm}
\includegraphics[keepaspectratio=true,scale=0.39]{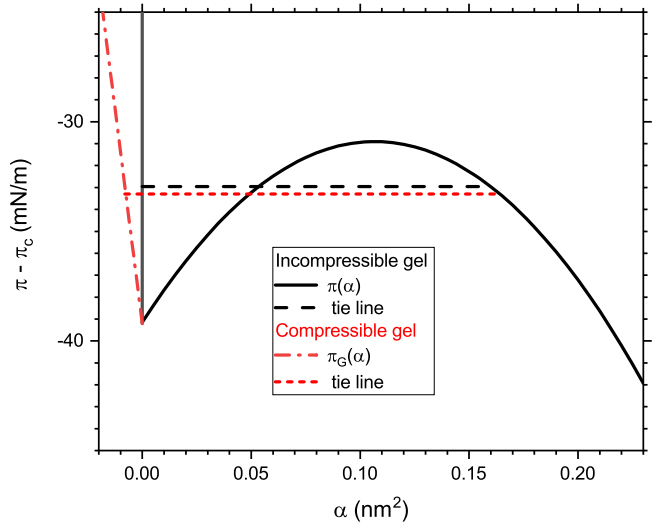}
\vspace{-0.2cm}
	\caption[]{Surface pressure vs. area/lipid isotherms for $t_1$ = -27.8 K, $B$ =  1.41 (mN/m)/deg, $C$ = 725 (mN/m)/nm$^4$, $D$ = 0.0038 nm$^2$/deg and $\pi_c$ = 33.0 mN/m (solid), with the tie-line (dashed).  A compressible gel phase is shown by the dash-dot line and the corresponding tie line by a short dash line.}
	\label{F3}
\end{center}
 \vspace{-0.5cm}
\end{figure}

Figure \ref{F3} shows the ${\pi}-A$ isotherm at $t_1$ = - 27.8 K for chosen values of the $B$, $C$ and $D$ parameters in Eq.~(\ref{eq4}).  The main transition occurs at $T_1$ = 24.0 $^\circ$C, so with Eq.~(\ref{eqt}) this choice gives $T_C$ = 51.8 $^\circ$C.   The usual Maxwell equal area construction that equates the free energies of the two phases then replaces the metastable and unstable portions of this isotherm with the horizontal tie-line at $\pi-\pi_C$ = - 33.0 mN/m. Since $\pi$ = 0 for a flaccid bilayer, this gives the critical pressure $\pi_c$ = 33.0 mN/m.   The increase in the experimental fluid phase area at the main transition is designated $\alpha_1$ and equals 0.16 nm$^2$. It is located at the end of the horizontal tie-line that is obtained from the Maxwell construction which requires exactly  
\begin{equation}
 t_1D = -2{\alpha}_1/3.
 \label{eq5}
 \end{equation}
 
 Of course, the gel phase is not totally incompressible. That could be taken into account by using a compressible gel phase line like what is shown in Fig.~\ref{F3}; for prominent visualization, it has been drawn to give a gel phase compressibility $K_A = -({\partial}\alpha/{\partial}{\pi}_t)/A$ that is 40\% as large as the fluid phase compressibility.  Even though that is an overestimate \cite{needham1988structure,lee2001all,tjornhammar2014reparameterized}, there is a  rather small difference in the corresponding tie line, so gel phase compressibility will be ignored henceforth.

\begin{figure}[ht]
\begin{center}
\advance\leftskip-2.5cm
\advance\rightskip-3cm
\vspace{-0.5cm}
\includegraphics[keepaspectratio=true,scale=0.39]{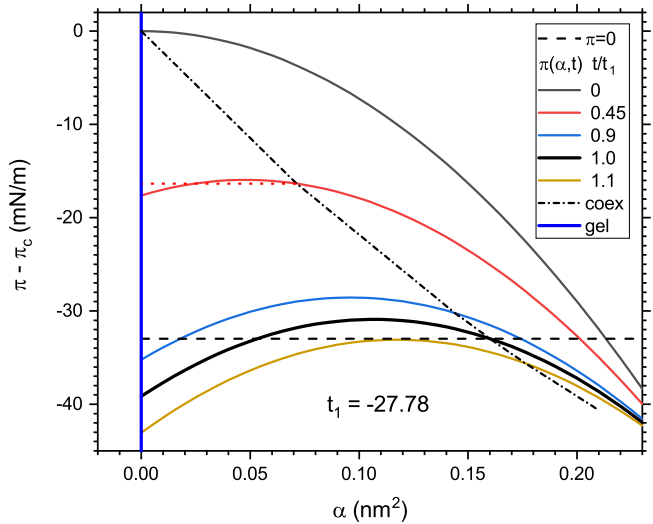}
\vspace{-0.5cm}
	\caption[]{Isotherms for the model in Fig.~\ref{F3} for some additional temperatures. The coexistence for $t$ = 0.45 $t_1$ is the red dotted tie line.  The dash-dot curve shows the locus of fluid phases that coexist with the gel phase at different temperatures and pressures.  The curve at the top is the critical isotherm. The black dashed line shows the experimentally accessible locus.}
	\label{F4}
 \end{center}
 \vspace{-0.5cm}
\end{figure}

As $t$ increases from $t_1$, the tie line in Fig.~\ref{F3} moves to experimentally inaccessible non-zero values of $\pi$ and its length becomes shorter and vanishes when $t$ = 0. This overall behavior is shown in Fig.~\ref{F4}. The point at $t$ = 0, $\alpha$ = 0 and $\pi$ = $\pi_c$ is a critical point with non-analytic thermodynamic properties.  As $t$ approaches 0, $-({\partial}\alpha/{\partial}t)_{\pi}$ diverges as $t^{-1/2}$ and the isothermal area compressibility $-({\partial}\alpha/{\partial}{\pi})_t/A$ diverges as $1/\alpha$ as $\alpha$ approaches 0.  

In the original toy model $\pi_c$ was zero.  However, the model was modified to allow for vacancies and that allowed for expansion in the lipid volume which was taken into account by adding an attractive van der Waals interaction as a mean field term. Along with head group and water interactions, positive values of $\pi_c$ were obtained and then the first order transition at $\pi$= 0 corresponds to the experimentally flaccid bilayer.  Straightforward experimental values of the interaction parameters resulted in reasonable agreement with experiment. That exact quantitative analysis is not repeated here.  The way that those prior results are taken into account in the present Landau type model is as justification for assigning the value of $\pi_c$ in Eq.~(4) that gives agreement with experiment when $\pi$ = 0 \cite{nagle1986theory}.

The values of the other parameters in Eq.~(\ref{eq4}) and in the caption of Fig.~\ref{F3} were chosen to obtain agreement with several types of experimental data. Here the appropriate thermodynamic equations are derived from Eq.~(\ref{eq4}).  The area compressibility modulus $K_A$ is obtained from Eq.~(\ref{eq4}) as

\begin{eqnarray}
K_A/A = - ({\partial}{\pi}/{\partial}{\alpha})_t = 2C({\alpha} + Dt) . 
\label{eq6}
 \end{eqnarray}
At the first order transition, Eq.~(\ref{eq5}) reduces this to
\begin{equation}
K_A/A_1 = (2/3)C\alpha_1   .
\label{eq7}
 \end{equation}
 from which the model parameter $C$ can be determined from experimental data for $A_1$, $\alpha_1$ and $K_{A1}$.  The equation of state (\ref{eq4}) also provides the change in area with temperature,
 \begin{eqnarray}
(\partial\alpha/\partial{t})_\pi = - \frac{(\partial\pi/\partial{t})_{\alpha}}{(\partial\pi/\partial\alpha))_t} = \frac{B - 2CD\alpha}{2C(\alpha + Dt)}    .
\label{eq8}
 \end{eqnarray}
which additionally involves both the $B$ and $D$ parameters.  At the first order transition, Eq.~(\ref{eq5}) reduces this to 
\begin{equation}
    (2/3)C\alpha_1(\partial\alpha/\partial{t})_{\pi} = B - 2CD\alpha_1  . 
\label{eq9}
 \end{equation}
Another independent relation is obtained from the enthalpy of the transition. First, the free energy $F_C$ is obtained by integrating $\pi = - (\partial{F}/\partial\alpha)_t$ to give
\begin{equation}
    F_C(\alpha,t) = - Bt\alpha  + (C/3)(\alpha^3+ 3Dt\alpha^2) - \alpha\pi_c  .    
\label{eq10}
\end{equation}
Entropy follows as
 \begin{equation}
   S_C(\alpha,t) = - (\partial{F}/\partial{t})_{\alpha} = B\alpha - CD\alpha^2  ,   
\label{eq11}
\end{equation}  
so the configurational entropy $S_C$ = 0 in the gel phase.  Then the first order transition enthalpy is 
\begin{equation}
 \Delta{H_1} = T_1\Delta{S_1} = T_1\alpha_1(B - CD\alpha_1). 
\label{eq12}
\end{equation} 
The three independent equations (\ref{eq7}), (\ref{eq11}) and (\ref{eq12}) enable determination of the $B$, $C$ and $D$ model parameters from experimental data.  Equation (\ref{eq5}) gives the value of $t_1$ and Eq.~(\ref{eqt}) gives the critical temperature $T_C$. 

The experimental value of $\Delta{H_1}$ for DMPC is 6.5 kcal/mole at $T_1$ = 297 K \cite{blume1983apparent}.  At $T_2$= 303 K the area $\alpha_2$ is 0.20 nm$^2$  \cite{koenig1997membrane,kuvcerka2005structure}. 
From an increase in the thickness of 0.013 nm \cite{chu2005anomalous} and a decrease of 1\% in the volume \cite{nagle1978lecithin}, the area at the main transition $A_1$ = 0.56 nm$^2$ and $\alpha_1$ = 0.16 nm$^2$ which is what is shown in Fig.~\ref{F2}. 
These give ($\partial\alpha/\partial{t})_{\pi}$ = 0.0067 nm$^2$/deg, somewhat larger than previous values (see p. 2634 of \cite{kuvcerka2005structure}). 
An additional reason to use a smaller value is the loss of one of the two lateral dimensions in the toy model that this Landau model is based on; since an area expansion is the square of a linear expansion, for small expansions this suggests a factor of ½ and the value 0.003 nm$^2$/deg is used for ($\partial\alpha/\partial{t})_{\pi}$. The experimental value of the area compressibility modulus $K_A$ at $T$ = 29 $^\circ$C is 234 mN/m \cite{rawicz2000effect}, but there are two factors that reduce this value when used in Eq.~(\ref{eq7}).  
First, the tilt independent bending modulus $K_C$ should also be smaller by about a factor of 0.6 \cite{nagle2017x} and this suggests that $K_A$ should also be smaller. Assuming as usual \cite{boal2012book,rawicz2000effect} that $K_C$ is proportional to $K_A$ times thickness squared and that the hydrocarbon chain thickness increases by 0.011 nm from T = 29 $^\circ$C to T = 24 $^\circ$C, an estimate of $K_A$ = 130 mN/m is used at $t_1$.  Second, it will be assumed that this value of $K_A$ should be further divided by a factor of three to take into account that each chain in the toy model only has two neighbors versus six neighbors in experiment.  Values of the ensuing model parameters are given in the caption to Fig.~\ref{F3}. 

The Gibbs free energy is obtained as
\begin{equation}
 G(t,\pi) = F(t,\alpha) + \pi A.  
\label{eq13}
\end{equation} 
It is properly concave because the specific heat is non-negative
\begin{equation}
C_{\pi} = T(\partial{S}/\partial{t})_{\pi} = 3(B - 2CD\alpha)^2/2C\alpha .  
\label{eq14}
\end{equation} 
Furthermore, the value of $C_{\pi}$ = 430 cal/mole/degree
is close to the experimental value of 370 cal/mole/degree \cite{wilkinson1982specific}.
 
\section{Tilt Free Energy Functional $F_\Theta$}
\label{sec3}

In this section a free energy $F_{\Theta}$ for the tilt degree of freedom is developed.  
For hydrocarbon chains tilted by angle $\theta$, following conventional notation \cite{chen1995phase,kozlov2000,andelman2008phase,raghu2012phase}, the tilt order parameter is written as $m$ = tan $\theta$.  Due to tilt symmetry, the free energy functional for tilt consists only of even powers of $m$, 
\begin{equation}
F_{\Theta}/A = \frac{1}{2}K_m m^2 + \frac{1}{4}b_4 m^4 + \frac{1}{6} b_6 m^6 + . . . ,   
\label{eq15}
\end{equation} 
where $K_m$ is the tilt modulus and $A$ is the area/lipid.  If one sets $K_m$ = $b_2t$ where $t$ remains the relative temperature, $t = T - T_C$, then this is analogous to the $\phi^4$ theory of magnetism when one terminates at the $m^4$ term with $b_4$ taken to be greater than 0 to ensure stability. Minimizing Eq.~(\ref{eq15}) with respect to $m$ yields $m^2$ = 0 for $t>$0, and for $t<$0 it yields a symmetry breaking spontaneous tilt $m^2$ = $-t/b_4$.  This $\phi^4$-like theory fails in that it predicts a critical point at $t$=0 with $K_m$ = 0 whereas DMPC has a first order transition at which $K_m\approx$ 20 mN/m is still non-zero \cite{nagle2017x}.  Of course, one can formally obtain a first order transition by add a cubic $b_3m^3$ term to Eq.~(\ref{eq15}), but this violates the symmetry between positive and negative tilting. 

Let us consider two ways to fix the preceding failure of the $m^4$ theory in Eq.~(\ref{eq15}). In this paragraph an ultimately unsuccessful, but illuminating, way is considered.  This way adds an $m^6$ term in Eq.~(\ref{eq15}) and assigns a negative value to $b_4$.  Adjustment of the parameters in this $m^6$ theory then provides a first order transition and a rather trivial way to reproduce the temperature dependence of the experimental tilt modulus by choosing $T_C$ = 291 K in Eq.~(\ref{eqt}).  Holding $b_2$ fixed then gives a value of $K_m$ twice as large at $T_2$ = 303 K as at the first order transition at $T_1$ = 297 K.   However, this $m^6$ theory fails because of the value that it predicts for the enthalpy of the transition 
\begin{equation}
\Delta{H}_{\Theta} = T_1\Delta{S}_{\Theta} = - T_1\Delta[(\partial{F_{\Theta}}/\partial{t})_{m}] = \frac{1}{2}\Delta[T_1b_2m^2] .  
\label{eq16}
\end{equation} 
Since $m$ = 0 in the fluid phase, this calculation needs only gel phase values, $A_G$ = 0.47 nm$^2$ and $\theta_G$ = 32° \cite{tristram2002structure} which gives $m^2$ = 0.39.  The value of $b_2$ is obtained as $K_m/t$ where $K_m$ = 20 mN/m and $t$ = 6 K at $T_1$ = 297 K.  The resulting $\Delta{H}_{\Theta}$ = 28 kcal/mole is four times larger than the total experimental enthalpy $\Delta{H}_{\Theta}$. It fails to include any contribution from trans-gauche isomerization and from the increase in van der Waals cohesive energy required for the volume increase at the transition. These latter two contributions have been estimated to account for nearly all the experimental $\Delta{H}$ \cite{nagle1980theory}. This $m^6$ theory is on the wrong track because it simply doesn’t account for the chain melting transition in lipid bilayers in other classes of lipids, like the phosphoethanolamines (PE), that have rather comparable transition quantities as the phosphocholines but have zero tilt in the low temperature phase \cite{mcintosh1980differences}.

In this paper, the $m^4$ free energy functional is modified in a different way that recognizes that the driver of the main phase transition is hydrocarbon chain melting.   It is then appropriate to couple the tilt free energy to the chain melting order parameter $\alpha$, so let us consider the following free energy functional $F_{\Theta}(m,\alpha)$ for the tilt contribution to the total free energy, 
\begin{equation}
F_{\Theta}(m,\alpha) =  \frac{1}{2}(g(\alpha) - b_2) m^2 + \frac{1}{4}b_4 m^4 
\label{eq17}
\end{equation}
where $b_2$ and $b_4$ are constant parameters. The major difference from the $m^4$ theory in Eq.(\ref{eq15}) is the removal of explicit temperature dependence and adding an area dependence in the function $g(\alpha)$ that is yet to be determined.  Setting $(\partial{F_{\Theta}(m,\alpha)/\partial{m}})_{\alpha}$ = 0 obtains potentially stable tilt values 
\begin{equation}
m^2 = (b_2 - g(\alpha))/b_4
\label{eq18}
\end{equation}
when $m^2$ is positive. Without loss of generality, let $g$(0) = 0 in the gel phase.  Then, the experimental value of $m^2$ = 0.39 in the gel phase \cite{tristram2002structure} provides the $b_2/b_4$ ratio and Eq.~(\ref{eq18})  verifies that $b_2$ is positive for the choice of its sign in Eq.~(\ref{eq17}).  For the fluid phase with $m$=0, the tilt modulus is
\begin{equation}
K_m(\alpha)=(\partial^2{F_{\Theta}(m,\alpha)/\partial{m}^2})_{\alpha}= g(\alpha) - b_2 .
\label{eq19}
\end{equation}
It goes negative for $\alpha$ = 0, as it should in order to break symmetry and induce the spontaneous tilt given by Eq.~(\ref{eq18}).  

Next, let us consider what is required of the free energy functional in Eq.~(\ref{eq17}). First, recall that a range of $(\alpha,m)$ is not stable thermodynamically when there is a first order transition in $\alpha$ just due to the $F_C$ term discussed in Section \ref{sec2}.  Nevertheless, that previous determination will be modified by $F_{\Theta}$ and that requires knowing the free energy functional in the unstable and metastable regions.  Second, recall that the reason there is spontaneous tilt in the gel phase is that the steric area of the lipid head groups $A_{head}$ determines the minimum area per lipid $A_G$. In contrast, the chain energy is minimized when the cross-sectional area is $A_0$.  The actual gel phase area $A_G$ is then the larger of $A_0$ and $A_{head}$.   When $A_0$ is smaller than $A_{head}$, for PC lipids but not for PE lipids, the cohesive van der Waals energy of the chains is minimized in the gel phase by cooperatively tilting by angle $\theta_G$ such that cos$\theta_G = A_0/A_{head}$ \cite{nagle1976theory,mcintosh1980differences,kheyfets2019origin}.  

We now apply this to $g(\theta)$ in Eq.~(\ref{eq19}).  As the constrained $\alpha$ is forced to increase from 0, the chain cross sectional area $A$ increases, so the chains tilt less and $m^2$ decreases.  This requires $g(\alpha)$ to increase with $\alpha$ in Eq.~(\ref{eq18}).  When $\alpha$ reaches the value 0.07 nm$^2$, at which $A = A_{head} = A_G$, the deepest cohesive chain energy is achieved when $m^2$ is zero. That requires $g$(0.07) = $b_2$ in Eq.(\ref{eq18}) and this also minimizes $F_{\Theta}$ in Eq.~(\ref{eq17}).   As $\alpha$ is increased further, $g(\alpha)$ further increases and $K_m(\alpha)$ in Eq.~(\ref{eq19}) increases from 0.  The first order transition is at $T_M := T_1$ = 24 $^\circ$C with $K_{m1}$ = 20 mN/m and it increases to $K_{m2}$ = 40 mN/m at $T_2$ = 30 $^\circ$C.  From the previous section $\alpha_2$ = 0.20 nm$^2$ and $\alpha_1$ = 0.16 nm$^2$. Then the values of $K_{m1}$ and $K_{m2}$ and Eq.~(\ref{eq19}) require  
\begin{equation}
g(0.20) - g(0.16) = g(0.16) - g(0.07) .
\label{eq20}
\end{equation}

To proceed further, it is necessary to choose a functional form for $g(\alpha)$.  A linear $g(\alpha)$ does not satisfy Eq.~(\ref{eq20}).  One could use a power series, but to minimize the number of additional parameters, $g(\alpha) = \Gamma\alpha^p$ is used.  Numerical fitting to Eq.~(\ref{eq20}) yields $p \approx$ 3 and then fitting to the $K_m$ values obtains $\Gamma$ = 5123 (mN/m)/nm$^6$ and $b_2$=0.94 mN/m. Finally, $b_4$ = $b_2$/0.39 = 2.41 mN/m follows from Eq.~(\ref{eq18}) for the gel phase with $g(\alpha)$ = 0 and the experimental $m^2$ = 0.39 value.
 
Now that all the parameters in Eq.~(\ref{eq17}) have been derived from experimental DMPC data, the final test is the magnitude of the transition enthalpy just due to the additional tilt term and ignoring the effect of tilt on the parameters in $F_C$.  Since enthalpy $H = F +TS + \pi{A}$,  the change in enthalpy at the transition just due to the tilting term is
\begin{equation}
\Delta{H}_{\Theta} = \Delta{F}_{\Theta} + T_M\Delta{S}_{\Theta}+{\pi}\Delta{A} = \Delta{F}_{\Theta} ,
\label{eq21}
\end{equation}
where the last equality comes because $\pi$ = 0 for flaccid bilayers and there is no explicit $T$ dependence in $F_{\Theta}(m,\alpha)$, so $S_{\Theta}$ = 0 in both phases.  In the fluid phase $F_m$ = 0 because $m^2$ = 0 and in the gel phase it equals -(1/4)$A_0b_2^2/b_4$.  This yields $\Delta{H}_{\Theta}$  = 0.01 kcal/mole which is quite small compared to the total experimental enthalpy of 6.5 kcal/mole. This is consistent with the greater number of degrees of freedom in chain melting compared to chain tilting. 

\section{Combining Tilt with Chain Melting }
\label{sec4}
\begin{figure}[ht]
\begin{center}
\advance\leftskip-2.5cm
\advance\rightskip-3cm
\vspace{-0.5cm}
\includegraphics[keepaspectratio=true,scale=0.39]{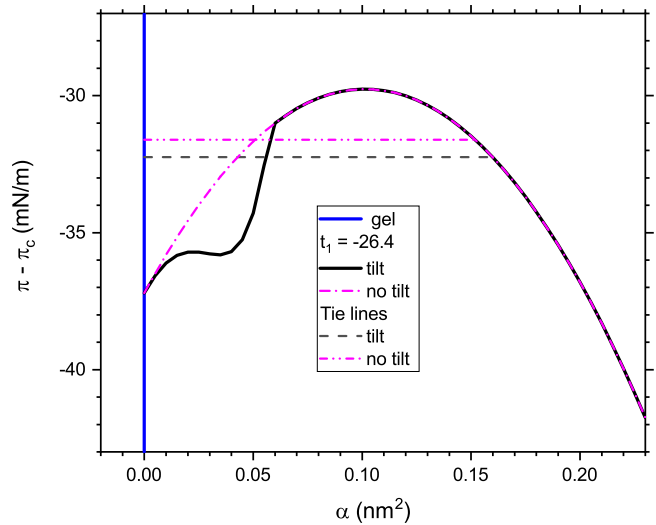}
\vspace{-0.5cm}
	\caption[]{Comparing isotherms at $t_1$ = -26.4 $^\circ$C with tilt (solid) and without tilt (dot-dashed) and tie lines with tilt (dashed) and without tilt (dash-dot-dot).}
	\label{F5}
 \end{center}
 \vspace{-0.5cm}
\end{figure}

The chain melting theory in Section \ref{sec2} took no consideration of the headgroup interaction that brings about tilt in the gel phase.  This section treats the effect of tilt on chain melting by combining the free energies from Sections \ref{sec2} and \ref{sec3} 
\begin{equation}
F_{C\Theta} = F_C + F_{\Theta}  .
\label{eq22}
\end{equation}
Then, a tilt pressure term must be added to the chain pressure shown in Fig.~\ref{F3}.  The tilt pressure is calculated as $-(\partial{F_{\Theta}(m,\alpha)/\partial{\alpha}})_m$ from Eq.~(\ref{eq17}), where $m$ is determined by Eq.~(\ref{eq18}) and is zero when $m^2$ would go negative according to Eq.~(\ref{eq18}).  The tilt pressure is negative as would be expected by adding another degree of freedom.  Although it is zero in the fluid phase where there is no net tilt, it affects the position of the tie-line, as seen in Fig.~\ref{F5}.  

\begin{figure}[ht]
\begin{center}
\advance\leftskip-2.5cm
\advance\rightskip-3cm
\vspace{-0.5cm}
\includegraphics[keepaspectratio=true,scale=0.37]{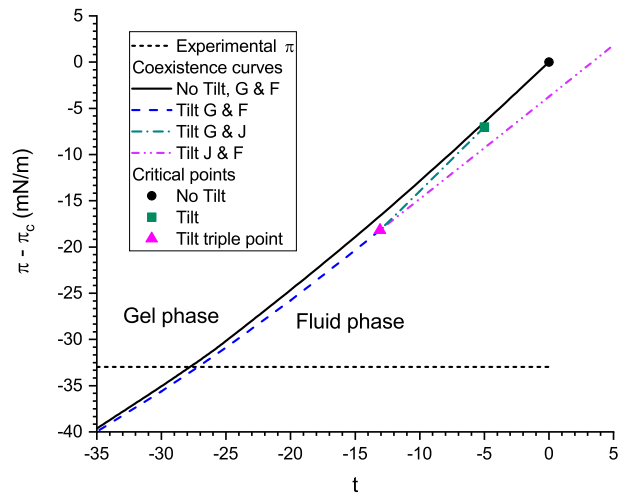}
\vspace{-0.5cm}
	\caption[]{The $\pi-t$ phase diagram showing the loci of the first order transitions and critical points with and without tilt.}
\label{F6}
\end{center}
\vspace{-0.5cm}
\end{figure}

Although adding tilt does not affect the first order transition very much, the phase behavior at higher temperatures and surface pressures is considerably affected because there is smaller variation of $\pi$ with $\alpha$ in the no-tilt isotherm whereas the tilt modification is explicitly temperature independent so it becomes more dominant at higher $t$.  Fig.~\ref{F6} shows the ensuing $\pi-t$ phase diagram with and without tilt.  The no-tilt phase line ends in a single critical point.  With tilt there is a triple point in Fig.~\ref{F6} with two first order lines extending to higher $\pi$ with a new intermediate phase between.  The upper line ends in a critical point like the no-tilt model.  The lower phase line extends to very high values of $\pi$. The appearance of two transitions as a function of temperature for values of $\pi$ above the triple point is suggestive of the lower and main transitions in DMPC and then the intermediate phase would be likened to the ripple phase. However, the differences in enthalpies and areas are far too small.  That the theory in this section ultimately misses getting both the main and the lower phase transitions is not surprising as there are more complex features to which we turn in the next section. 

\section{Review of the Ripple Phase}
\label{sec5}
\begin{figure}[ht]
\begin{center}
\advance\leftskip-3cm
\advance\rightskip-3cm
\includegraphics[keepaspectratio=true,scale=0.37]{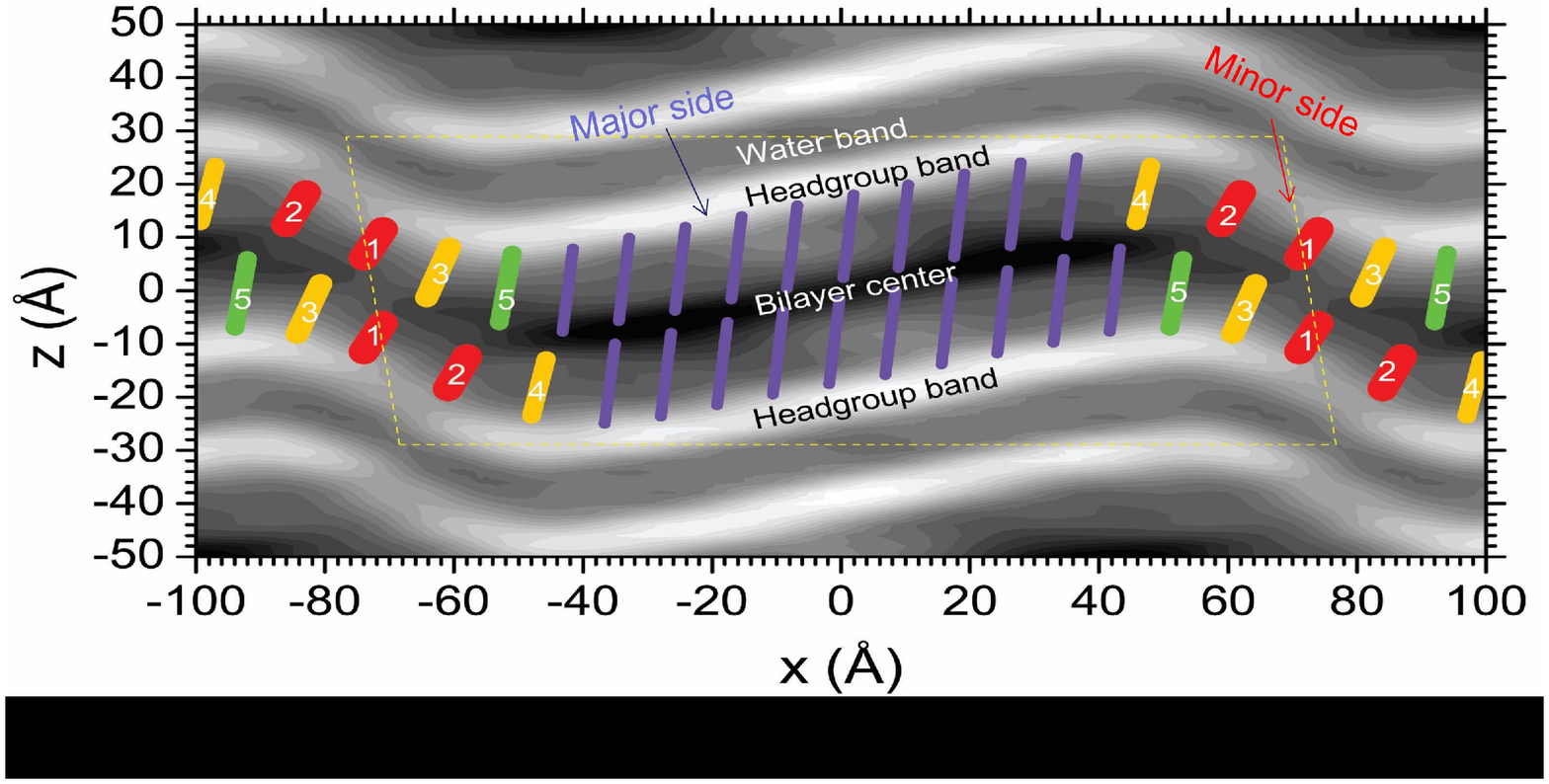}
\vspace{-0.3cm}
\caption[]{Structure of the DMPC ripple phase adapted from \cite{akabori2015structure}. The sample was a stack of bilayers at $T$ = 18 $^\circ$C.  Grey scale shows the electron density which is highest in the headgroup band and lowest in the bilayer center. Coarse grained representations of chain conformations are superimposed in color.  The unit cell is shown by yellow dashed lines.  The upward-sloping major side of the ripple is in the center of the unit cell and the minor side is at the edges.}
\label{F7}
\end{center}
\end{figure}

Although there are thermal out-of-plane fluctuations, especially in the fluid phase, the time averaged bilayer is flat, in both the gel and fluid phases, as has been assumed in the preceding theory.  In contrast, the ripple phase breaks the flat symmetry by having static out-of-plane structure that is singly periodic in one of the in-plane directions.  The most recent high resolution x-ray study obtained an electron density profile that is shown in Fig.~\ref{F7}.  As had originally been recognized \cite{tardieu1973structure}, the profile is asymmetric with a major, upward-sloping, longer side and an even more downward-sloping, shorter minor side.  The electron density in the headgroup region is primarily due to the electron dense phosphate headgroups so the higher electron density in the major side headgroup band means a smaller area per lipid compared to the minor side with its lower electron density.  Fig.~\ref{F7} also superimposes chain conformations obtained from wide angle x-ray scattering on the electron density profile. The gel-like chains in the major side are caricatured as elongated and thin. In the minor side the chains are portrayed as shorter and more fluid-like on average, with more distance between them consistent with the lower electron density in the minor side headgroup region.

The height profile $z(x)$ of the ripple is quantified in Fig.~8.  Also shown is the area profile $\alpha(x)$ that is obtained by smoothing the electron density data from Fig.~\ref{F6} in \cite{akabori2015structure}.  Note that $\alpha(x)$ = 0.049 in the major side is greater than zero because the chains are less tilted by $\theta_{tilt}$ = 18$^\circ$ relative to the local bilayer normal compared to 32$^\circ$ in the gel phase. A smaller tilt in the ripple phase has also been reported from infrared spectroscopy data \cite{leBihan1998study}.  It is also estimated from \cite{akabori2015structure} that the maximum $\alpha(x)$ = 0.15 nm$^2$ in the center of the minor side between chains designated as 1 and 2 in Fig.~\ref{F7}.  It may also be reiterated \cite{akabori2015structure} that the relative offset in $x$ of the locations of the monolayer minimal headgroup electron densities weighs strongly against interdigitation of chains in the minor side.  Obtaining the structure of the ripple phase continues to be a challenge for simulations \cite{deVries2005molecular,kranenburg2005phase,lenz2007structure,Gezelter2008dipolar,khakbaz2018investigation,karttunen2023elucidating}.
\begin{figure}[ht]
\begin{center}
\advance\leftskip-2.5cm
\advance\rightskip-3cm
\vspace{-0.5cm}
\includegraphics[keepaspectratio=true,scale=0.37]{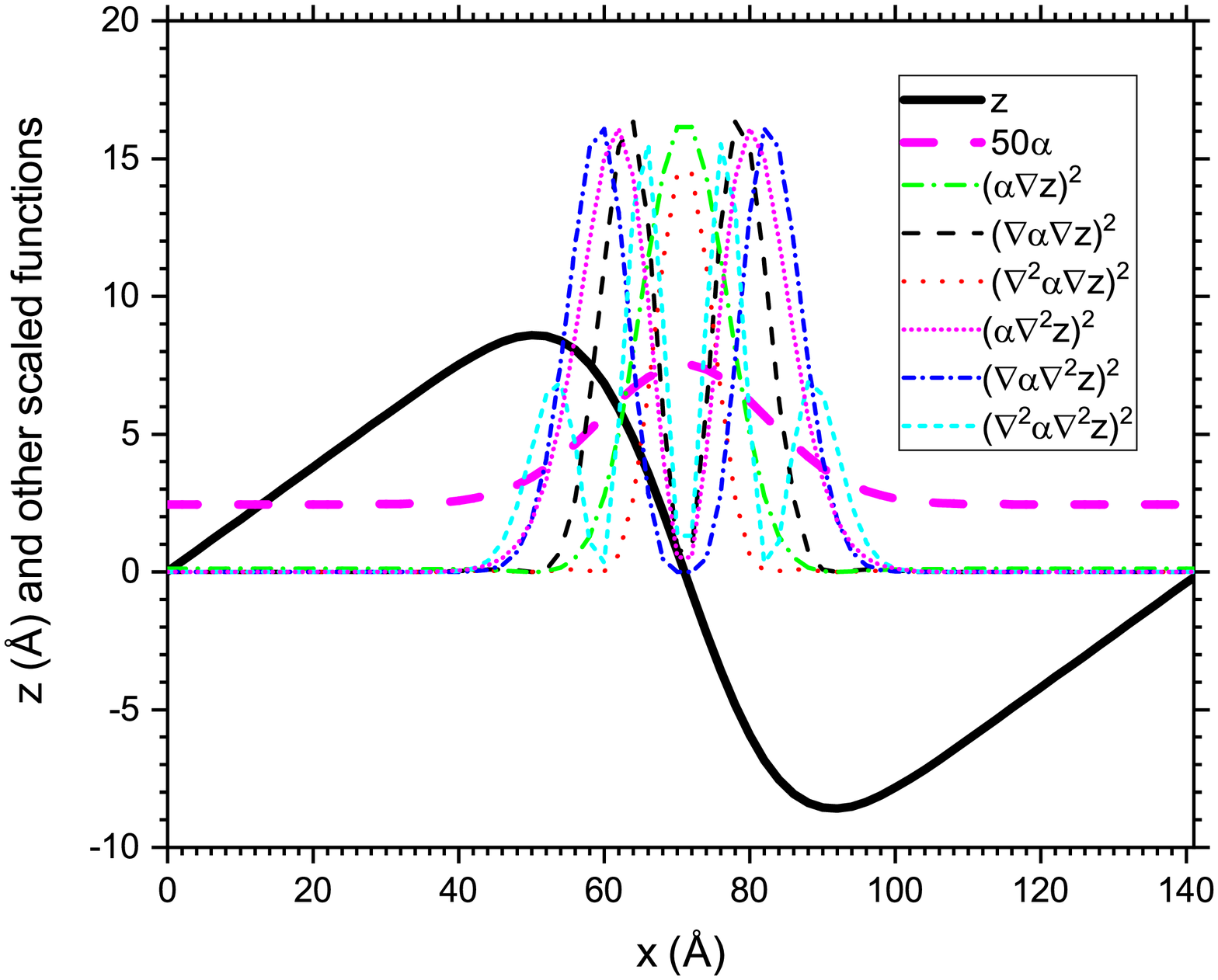}
\vspace{-0.5cm}
\caption[]{The thick black line shows the ripple phase height profile 
$z(x)$ of the headgroup band of one monolayer in Fig.~\ref{F7}.  The thick dashed magenta line shows the corresponding area profile $\alpha(x)$ times 50.  The broken lines show six potential additions that could account for a heterogeneous coupling term; they are arbitrarily scaled for visibility and the functional forms are identified in the legend.}
\label{F8}
\end{center}
\vspace{-0.75cm}
\end{figure}

\section{Two Phase Transitions}
\label{sec6}
\begin{figure}[ht]
\begin{center}
\advance\leftskip-3cm
\advance\rightskip-3cm
\vspace{-0.5cm}
\includegraphics[keepaspectratio=true,scale=0.39]{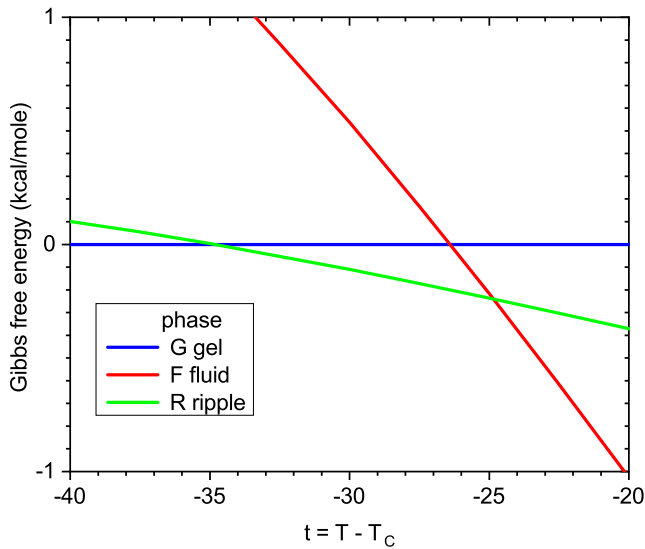}
\vspace{-0.5cm}
\caption{Model Gibbs free energies that give a ripple phase over a ten degree interval with $T_C$ = 48.9 $^\circ$C, $T_M$ = 24.0 $^\circ$C and $T_L$ = 14.0 $^\circ$C.}
\label{F9}
\end{center}
\vspace{-0.5cm}
\end{figure}

To address the phase transitions further, consider the Gibbs free energies, $G_G$ for the gel phase, $G_F$ for the fluid phase, and $G_R$ for the ripple phase, as functions of temperature.  For the experimental trajectory $\pi = 0$, $G$ is the same as the Helmholtz free energy $F$. In Fig.~\ref{F9} the free energy of the gel phase has been simplified to be 0 at all temperatures, thereby ignoring higher order contributions like thermal expansion of the chain packing \cite{sun1994order}. 
For the ripple phase, the simple approximation is made that the temperature dependence of $G_R$ is a linear combination of a gel-like major side and a fluid-like minor side as well as a new term $G_H$ that depends on heterogeneity.
\begin{equation}
G_R(T) = \gamma{G_G(T)} + (1-\gamma)G_F(T) + G_H.
\label{eq23}
\end{equation}
In first approximation, $G_H$ will be considered to be temperature independent.
Accordingly, the slope of $G_R(T)$ lies between those of $G_G(T)$ and $G_{F}(T)$. Importantly, if $G_H <$ 0, then there will be two transitions as shown in Fig.~\ref{F9}.

Since transition enthalpy 
\begin{equation}
\Delta{H} = T_1\Delta{S} = -T_1\Delta(\partial{G}/\partial{T})_{\pi},
\label{eq24}
\end{equation}
 the value of $\gamma$ in Eq.~(\ref{eq24}) determines the transition enthalpy of both the lower transition $\Delta{H_L}$ and the main transition $\Delta{H_M}$.  Because the specific heat is quite small compared to the transition enthalpies, $G_F(T)$ has nearly constant slope, so $\Delta{H_M}/\Delta{H_L} \approx \gamma/(1-\gamma)$. Since the experimental $\Delta{H_M}/\Delta{H_L}$ is about 5 \cite{blume1983apparent}, Eq.~(\ref{eq24}) assigns $\gamma\approx$ 5/6 of the ripple thermodynamics to the major side. That suggests a relatively larger major side fraction $\gamma$ than visualized in Figs. \ref{F7} and \ref{F8}.  However, this also assigns 1/6 of the ripple to a pure fluid minor side, and it is clear from Fig.~\ref{F7} that the minor side is more ordered on average than the pure fluid phase, so the $\gamma$ value that agrees with experimental values of the transition enthalpies is reasonable.  Finally, the difference in experimental transition temperatures determines the value of $G_H$ in Eq.~(\ref{eq23}). However, note that $G_H$ will have to be more negative if $G_G$ in Eq.~(\ref{eq23}) is replaced by a positive value to account for the smaller $\theta_{tilt}$ in the major side compared to the gel phase that is noted in the previous section.  Also, note that the experimental specific heat \cite{wilkinson1982specific,riske2009lipid} and the thermal rate of volume expansion \cite{nagle1978lecithin} are greater in the ripple phase than in the $G$ and $F$ phases, so $G_R(T)$ should be more concave than allowed by Eq.~(\ref{eq23}), which assumes that $G_H$ is independent of temperature. Also, the amplitude of the ripple has been reported to increase as temperature increases \cite{zasa1996amplitude,cunningham1998ripple}, so adding temperature dependence to $G_H$ would allow this simple model to be more realistic, but structural data of comparable quality to Fig.~7 are not available to pursue this.

\section{Discussion}
\label{sec7}
Chain melting is the most important thermodynamic driver of the main phase transition \cite{nagle1980theory}. Similar to much of the literature, Section \ref{sec2} treats this with a continuum free energy functional involving an order parameter which is here taken to be chain area $\alpha$ rather than the essentially equivalent bilayer thickness used by others \cite{marder1984theory,goldstein1988model,andelman2008phase,raghu2012phase}. More importantly, the functional form adopted in this paper differs from the conventional one to better accommodate the steric interactions that account for a larger area compressibility modulus in the gel phase than what the conventional form provides.  This functional form comes from a detailed model of sterically hindered chain packing that has a 3/2-order critical point \cite{nagle1975critical,nagle1980theory} rather than from the conventional $\phi^4$ form appropriate for soft spin-type interactions. 

Chain tilt is an important secondary order parameter for lipid bilayers that have large headgroups that force tilt in the gel phase.  Although the functional form that is used in Section \ref{sec3} is similar to the $\phi^4$ form, it differs by coupling to the chain area $\alpha$ and its temperature dependence rather than to temperature directly.  This treatment quantitively reproduces the recently observed temperature dependence of the tilt modulus data above the main transition \cite{nagle2017x}.  This decrease in the tilt modulus as temperature is lowered to the main transition is the best experimental evidence thus far for a critical point in lipid bilayers.  The theory predicts that the observed first order transition would become critical if the lateral pressure $\pi$ could be increased sufficiently, but it has not yet been possible to do that experimentally. 

Chain melting and chain tilting together provide a fundamental understanding of the main transition at a qualitative level, and the theory in Section \ref{sec4} provides quantitative support.  However, this leaves unexplained the lower transition and the ripple phase.  It has been recognized in the many papers on the subject that this is an interesting theoretical challenge \cite{doniach1979thermodynamic,marder1984theory,carlson1987theory,goldstein1988model,chen1995phase,mackintosh1997internal,hansen1998fluid,andelman2008phase,raghu2012phase,hawton1986van,scott1991theories,cevc1991polymorphism,heimburg2000model,Nelson1996role,misbah1998transition}. At the continuum level it has long been recognized that at least one heterogeneous Ginzburg-like term is required in the free energy to obtain a phase that is not spatially uniform \cite{marder1984theory,goldstein1988model,chen1995phase,mackintosh1997internal,andelman2008phase,raghu2012phase}.  Such theories posit one or more order parameters and then consider terms that involve their gradients and divergences to lowest order.  The latest example considered many such terms, also with two order parameters \cite{raghu2012phase}. To obtain a modulation profile $z(x)$, a spatial functional form with two sinusoidal terms was assumed and the parameters in this spatial form were then determined to minimize the free energy which had its own parameters.  These latter parameters were then varied to obtain spatial modulation of the height profile that appears similar to the experimental data, but their main order parameter $\psi$ is essentially sinusoidal instead of being constant in the major side of the ripple. 

Compared to the approach \cite{raghu2012phase} in the previous paragraph, Section \ref{sec6} simply takes the experimental height profile as given, thereby avoiding having to assume a spatial functional form with its undetermined parameters.  There are again many possible heterogeneous terms (see the legend in Fig.~8) that could be added to the free energy to provide a negative value of $G_H$ in Eq.~(24) that then gives a ripple phase and a lower transition.  Although this obtains suitable agreement with experiment, it does not discriminate between these possible heterogeneous terms.  More unsatisfyingly, the development in Section \ref{sec6} shares with all the continuum theories of the ripple phase that such terms are quite phenomenological, lacking underlying physical insight into the interactions of lipid molecules that could account for them. 

In contrast to our physical understanding of why there should be a transition from a tilted gel phase to the fluid phase, it is unclear to this author that there is even qualitative understanding of what it is at the molecular level that brings about the ripple phase and the lower transition. An important objective is to find a physical criterion that limits the size of the major side, and a new qualitative suggestion has been made regarding kink-block structures in the discussion in \cite{akabori2015structure}.  Previous theories that focus on this objective have involved splayed domains \cite{carlson1987theory} and next nearest neighbor interactions \cite{hawton1986van,scott1991theories}, but these, along with other notable theories \cite{doniach1979thermodynamic,cevc1991polymorphism,heimburg2000model} provided ensuing ripple structures that differ considerably from the ripple structure in Fig.~\ref{F7}.  It could be insightful if theories involving fundamental interactions could discriminate between the different continuum heterogeneous forms that are mentioned in the legend to Fig.~\ref{F8} but it is beyond the scope of this paper to attempt such connections. 

It should also be noted that most theories, including the one in this paper, assume that it is sufficient to assign order parameters just to the bilayer, but the experimental structure in Fig.~\ref{F7} suggests that one might have to consider an order parameter for each monolayer with coupling between monolayers as proposed in \cite{Nelson1996role,hansen1998fluid}.  Fig.~\ref{F7} also emphasizes that the sample was a stack of closely spaced bilayers and that raises the issue of whether interactions between bilayers that have only been considered by a few theories \cite{cevc1981undulations,goldstein1989structural,misbah1998transition} might be essential for formation of the ripple phase and a lower transition.  There are reports that uni-lamellar vesicles (ULVs) do not have a lower transition \cite{Mouritsen2010interlamellar,Almeida2015guvs}, while earlier papers did report a calorimetric pretransition, although much attenuated \cite{parente1984phase,biltonen1993use,mason1999small}.  Visualizations of ripples have been reported in ULVs \cite{mcconnell1980rippled} and also in mica supported double bilayers \cite{mouritsen2003temperature} and the top layer on a stack of bilayers \cite{schafer2008atomic}.  Although interbilayer and intermonolayer interactions may be important for the detailed structure of the ripple phase, the theory in this paper assumes, along with most other theories, that a single bilayer model remains relevant, especially for the main phase transition whose enthalpy is adequately accounted for by chain melting \cite{nagle1980theory}.

Even though the particular continuum theory in this paper does not provide the desired fundamental understanding of what causes the ripple phase and the lower transition beyond invoking heterogeneous terms in a continuum model, it successfully accommodates a great deal of experimental data, more than previous continuum theories \cite{doniach1979thermodynamic,marder1984theory,carlson1987theory,goldstein1988model,chen1995phase,mackintosh1997internal,hansen1998fluid,andelman2008phase,raghu2012phase} that also did not agree nearly so well with the more recent structure in Fig.~\ref{F7}.  Finally, this is the first and only attempt to date to account theoretically for the relatively recently observed critical-like behavior of the tilt modulus \cite{nagle2017x}.

Acknowledgements:  The author thanks Dr. Saheli Mitra for comments on the manuscript. 

\section{Appendix}
Proof is given of the statement in the text that the $\phi^4$ theory requires 
\begin{equation}
K_{AG}/A_G = K_{AF}/A_F .
\label{a1}
\end{equation}
The area modulus $K_A$ is given by
\begin{equation}
K_A/A := - (\partial\pi/\partial\alpha)_t = (\partial^2F/\partial\alpha^2)_t,
\label{a2}
\end{equation}
so Eq.~\ref{a1} follows if the second derivatives of $F$ are equal for the gel $G$ and the fluid $F$ phases at the main transition temperature $t_1$ and at their respective areas, 0 for the gel phase and $\alpha_1$ for the fluid phase.  
For both phases $F$ and $(\partial{F}/\partial\alpha)_t$ equal $0$. Together these require $\alpha_1 = -2b_3/3b_4$ and $b_2t_1 = 2b_3^2/9b_4$. The second derivative,  
\begin{equation}
\partial^2F/\partial\alpha^2)_t = b_2t + \alpha(2b_3 + 3b_4\alpha),
    \label{a3}
\end{equation}
has the same value, $b_2t$, in the gel phase because $\alpha_G = 0$, and in the fluid phase because $\alpha_F = \alpha_1 = -2b_3/3b_4$.  QED
\section*{References}
\bibliography{refs}

\end{document}